\documentstyle[12pt]{article}
\def\ni{\noindent}

\newcommand{\mk}{|{\bf k}|}
\begin{document}
\begin{tabbing}
\`SUNY-NTG-93-21\\
\`May 1993
\end{tabbing}
\vbox to  0.8in{}
\centerline{\Large \bf Low-mass dileptons from nonequilibrium QGP}
\vskip 2.5cm
\centerline{\large A.  Makhlin }
\vskip .3cm
\centerline{Department of Physics}
\centerline{State University of New York at
Stony Brook}
\centerline{Stony Brook, New York 11794}
\vskip 0.35in
\centerline{\bf Abstract}
 The rate of the emission of the high energy low-mass dileptons from the QGP is
found in the first nonvanishing order with respect to strong coupling.
We base on the real-time kinetic approach [2] without
an explicit assumption about a complete thermal
equilibrium in the emitting system.  For the class of the partons
distributions which may simulate that of the "hot glue scenario"[1]
the rate of emission is found analytically .

\vfil
\noindent
${^\dagger}$ E-mail address: "makhlin@sbnuc"
\eject

\pagebreak

\vskip 2.5cm

{\bf 1. } Most of the results concerning the rate of
emission of low-mass dileptons
from quark-gluon plasma implicitly assumed that the emitting
plasma is in thermal equilibrium. More general consideration
is needed in connection with some modern scenario of the QGP formation which
predict not complete thermalization of QGP but suppression of the quark
component against the background of a hot glue [1].
For this reason we can not use any of standard thermal field approaches
and base on the field kinetic technique developed in paper [2].

{\bf 2.}Contrary to other electromagnetic probes of the QGP as heavy dileptons
and real photons,
the very notion of a low-mass dileptons should be carefully specified.
It is not enough to chose the kinematic region as $E>>M,T$ where the
energy $E$ of the dilepton is much larger than its invariant mass
$M$ and the temperature $T$. Emission of the virtual photon from
the hot nuclear matter is a complicate dynamical phenomenon. The inclusive rate
of dilepton emission includes many exclusive channels. Some of them
have a threshold behavior. For instance, the first Born's term.   For the
Drell-Yan prototype of the dilepton emission the threshold is defined by the
minimum $2m_{q}$ of the invariant mass of the annihilating particles.
Indeed,  the rate of the dilepton emission is given by[2]
\begin{equation}
 {{dN_{e^{+}e^{-}}}\over{d^{4} k d^{4}x}}=
 {{ie_{0}^{2}}\over{2(2\pi)^{6}}}{\pi \over M^{2}}
 \sqrt{1-{4m_{l}^{2} \over M^{2}}}[1-{1\over 3}(1-{4m_{l}^{2}\over M^{2}})]
\theta(M^2-4m_{l}^{2}) g_{\mu\nu} \Pi^{\mu\nu}_{10}(-k)
\end{equation}
where $m_{l}$  is a lepton mass.
If we denote  $g_{\mu\nu} \Pi^{\mu\nu}_{10}(-k) $ as $\pi(k)$ and confine
calculations to the lowest perturbation order we easily get
\begin{equation}
 \pi_{Born}(k)=- {{3ie^{2}}\over{4\pi}}(1+{2m^{2} \over M^{2}})
 \sqrt{1-{4m^{2} \over M^{2}}} M^{2}e^{-ku/T}\theta(M^2-4m^2)
\end{equation}
The square root here is typical for many thresholds in quantum mechanics.
If $M^{2}<4m^{2}$ then initial quarks are too heavy and the emission of the
dilepton is possible only due to the processes of the next perturbative orders.
How shall we proceed if we want to deal only with small masses: low-mass
virtual photons {\it and} massless quarks? For real photons the answer is
evident. Two polarizations of a real photon are consistent only with $M=0$
and the Born's term is absent.

{\bf 3.}To find out if the Born's term is present in the dilepton production by
the
massless quarks we are to compare  the dilepton mass with some scale inherent
to the emitting system.

 In the previous papers [3,4] we considered emission of
the heavy dileptons and photons and
 found that the zero quark masses are potentially dangerous only in
 collinear geometry of the emission process. Indeed, for the hard quarks
 the influence of the  rest mass onto the energy balance is negligible.
 The finite mass works  as a geometric parameter, the Compton length
 $l_{c}=1/m$, thus restricting the  domain  of the interaction.
 In the media  this  role is played by the minimal length $l_{fs}$ defined
 by the amplitude of the forward scattering.
 If we demand that $E>>M,T$ then the initial quarks should be really
 hard and the estimate   $l_{fs}\sim 1/gT \sim 1/m_{therm}$ is very reliable.
 A potential danger  hidden in the collinear geometry of annihilation
 emerges from the low relative momentum of the $q\bar{q}$-pair because
 in this case the Coulomb interaction is not a perturbative effect.
 But it does not seem to be important in neutral plasma at high temperatures.

  So the boundary between low-mass and heavy dileptons lies at
 $M \sim 2m_{therm}$. Above this threshold the Born's term (the
 direct $q\bar{q}$-annihilation) very quickly grows up and dominates over
 the next order contributions until $M/T$ reaches value about 10 [3].

 {\bf 4. } The threshold $k^{2}=4m^{2}$ of
 the Born's term is that for the one-loop
 radiative corrections. Below it they have nothing to interfere with. So
 only "real" processes contribute the rate of emission. In this case the trace
 of the electromagnetic polarization tensor  reads as [3]

 \begin{eqnarray}
\pi_{real}=-{ie^{2}g^{2}N_{c}C_{F} \over  2\pi^{5} } \int d^{4}pd^{4}q
\delta(q^{2})\delta[(p-k)^{2}-m^{2}]\delta[(p+q)^{2}]
 (SW_{a}+SW_{c})  \times \nonumber\\
\times \{1+ {A \over (p^{2}-m^{2})^{2}}+{B \over p^{2}-m^{2}}  +
{ C \over p^{2}-m^{2}+2kq}   \}
\end{eqnarray}
where  we denoted:
\begin{eqnarray}
A=2m^{2}(k^{2}+2m^{2}),\;\;\;\;\;
B=3m^{2}+k^{2}+2(kq)-{ 4m^{4}-(k^{2})^{2} \over 2(kq)} \nonumber \\
C=(k^{2}+m^{2})+{ 4m^{4}-(k^{2})^{2} \over 2(kq)}
\end{eqnarray}
The expression in the curly  brackets is
a  sum  of the squared moduli of the matrix
elements of the annihilation process, $q\bar{q}\rightarrow g\gamma$, or
Compton process, $qg\rightarrow q\gamma$ and
$\bar{q}g\rightarrow \bar{q}\gamma$.
Specification of the process is due to statistical weights $SW_{a,c}$.

  As in the previous papers [2,3,4]  we will not assume
that quarks and gluons are in thermal equilibrium. Instead we shall use the
Boltzmann-like distributions with fugacities
$\zeta_{Q}$ and $\zeta_{G}$ which all together will be considered as the
measure
of the chemical and kinetic nonequilibrium in the
quark-gluon system. Then statistical weight of the
annihilation process with the emission of a gluon looks as
\begin{eqnarray}
SW_{em}=\theta(k_{0}-p_{0}) \theta(q_{0}+p_{0}) \theta(q_{0})
n_{F}(k_{0}-p_{0})n_{F}(q_{0}+p_{0})[1+n_{B}(q_{0})] \approx \nonumber \\
\approx \theta(k_{0}-p_{0}) \theta(q_{0}+p_{0}) \theta(q_{0})
\zeta_{Q}^{2} e^{-(ku)/T}  (e^{-(qu)/T}+\zeta_{G}e^{-2(qu)/T})
\end{eqnarray}

For the Compton rate of the dilepton emission the statistical weight equals to
\begin{eqnarray}
SW_{com}=-\theta(p_{0}+q_{0}) \theta(p_{0}-k_{0}) \theta(-q_{0})
n_{F}(p_{0}+q_{0})[1-n_{F}(p_{0}-k_{0})]n_{B}(-q_{0})-   \nonumber \\
-\theta(-p_{0}-q_{0}) \theta(k_{0}-p_{0}) \theta(-q_{0})
n_{F}(k_{0}-p_{0})[1-n_{F}(-p_{0}-q_{0})]n_{B}(-q_{0})- \approx \nonumber \\
\approx  -\zeta_{Q} \zeta_{G} \{ \theta(p_{0}+q_{0}) \theta(p_{0}-k_{0})
\theta(-q_{0}) e^{-pu/T}[1-\zeta{Q} e^{-(pu-ku)/T}] + \nonumber \\
+\theta(-p_{0}-q_{0}) \theta(k_{0}-p_{0}) \theta(-q_{0})
e^{-ku/T} e^{(qu+pu)/T}[1-\zeta_{Q} e^{(qu+pu)/T}] \}
\end{eqnarray}

Then we can perform an exact integration over $p$ using the Breit reference
system where ${\bf k + q}=0$, c.m.s. of the reaction $q\bar{q}\rightarrow
g\gamma$ This immediately leads to
\begin{equation}
\pi_{ann}=-{ie^{2}g^{2}N_{c}C_{F} \over  4\pi^{4} }
\zeta_{Q}^{2} e^{-ku/T} \int d^{4}q
\delta(q^{2}) \theta(q_{0}) \theta[(k+q)^{2}-4m^{2}]
\{e^{qu/T}+ \xi_{G} e^{-2qu/T} \}{\cal F}_{a}(kq;M^{2})
\end{equation}

 where

\begin{eqnarray}
{\cal F}_{a}(x;M^{2})= (1+ { M^{2}+2m^{2} \over x} + {M^{4}-4m^{4}
\over 2x^{2} }) \ln{ 1- \sqrt{1-4m^{2}/(M^{2}+2x)}
\over  1+ \sqrt{1-4m^{2}/(M^{2}+2x)} } \nonumber \\
+(1+ { M^{2}+2m^{2} \over x} + {M^{2}(M^{2}+2m^{2})
\over 2x^{2} }) \sqrt{1-{4m^{2} \over M^{2}+2x} }
\end{eqnarray}

For the Compton rates we start with the chain of changes of variables,
 $p\rightarrow -p+k-q$,
in the first term and $q \rightarrow -q-p$ in both terms we may also easily
perform the  integration over $p$  using the same Breit reference system.
It gives
\begin{equation}
\pi_{compt}=-{ie^{2}g^{2}N_{c}C_{F} \over  4\pi^{4} }
\zeta_{Q}\zeta_{G} e^{-ku/T}
\int d^{4}q \delta(q^{2}-m^{2}) \theta(q_{0})
[e^{-qu/T}-\zeta_{Q}e^{-2qu/T}]{\cal F}_{c}(kq;M^{2})
\end{equation}
where q is the momentum of the (anti)quark in the final state and
\begin{eqnarray}
{\cal F}_{c}(x;M^{2})= - { 4m^{2}+M^{2}-2x-2(M^{4}-4m^{4})/(M^{2}+2x)
\over 2 \sqrt{x^{2}-M^{2}m^{2}} } \ln{ m^{2}+x+ \sqrt{x^{2}-M^{2}m^{2}}
\over  m^{2}+x- \sqrt{x^{2}-M^{2}m^{2}}  } \nonumber \\
+{4(M^{2}+2m^{2}) \over M^{2}+2x }+{(M^{2}+2x)(m^{2}+x) \over
(M^{2}+m^{2}+2x)^{2} }
\end{eqnarray}

{\bf 5.} Further integration, first angular, is easily performed
 in the rest frame of the dilepton were ${\bf k}=0$.
 Having expressed the result in the invariant form and using
 $(k_{0},{\bf k})$ as the components of 4-vector $k$ in the rest frame of the
 emitting media we get
\begin{equation}
\pi_{a}={ie^{2}g^{2}N_{c}C_{F} \over  2\pi^{3} }
\zeta_{Q}^{2} e^{-ku/T} M^{2}
\int_{(4m^{2}/M^{2}-1)/2}^{\infty} [S(k,y)
+\zeta_{G}  S(k,2y)] {\cal F}_{a}(M^{2}y)ydy
\end{equation}
The nonzero lower limit has appeared because a dilepton production without
emission of a gluon is impossible in this  kinematic region.  In sequence,
we need do radiative corrections to eliminate the IR singularity at zero
momentum of the emitted gluon.
The terms with the extra factor $\zeta_{G}$ relate to the induced emission of a
gluon and we denoted,
\begin{equation}
  S(k,y)=  e^{-k_{0}y/T}{\sinh(\mk y/T) \over (\mk y/T)}
\end{equation}
The term with the extra factor $\xi_{G}$ relate to the induced emission of a
gluon.

If we are interested only in the distribution of the dileptons over their
invariant masses we are to follow Eq. (1) and get function
\begin{eqnarray}
U_{a}(y)= \int { d^{3} {\bf k} \over 2k_{0} } e^{-k_{0}/T}S(k,y)=
2\pi M^{2} { K_{1}(M \sqrt{1+2y}/T) \over  M\sqrt{1+2y}/T } \approx \nonumber
\\
\approx 2^{-1} (2\pi)^{3/2} M^{1/2} T^{3/2}
(1+2y)^{-3/4} e^{-M\sqrt{1+2y}/T}
\end{eqnarray}
It can be first integrated over the history of the system (if necessary,
with fugacities $\zeta$) and only then substituted into the integrands
of Eq.(11).

For the Compton rate, proceeding as above, we get
\begin{equation}
\pi_{compt}=-{ie^{2}g^{2}N_{c}C_{F} \over  2\pi^{3} }
\zeta_{Q} \zeta_{G} e^{-ku/T} m^{2} \int_{0}^{\infty} {x^{2}dx \over  x_{0}}
{\cal F}_{c}(x) [C(x,k,T)-\zeta_{Q}C(x,k,{T \over 2})]
\end{equation}
where  $x_{0}=\sqrt{1+x^{2}}$ and
\begin{equation}
  C(x,k,T)=  e^{{-k_{0}m \over MT}x_{0}}
  {\sinh({\mk m \over MT}x)  \over ({\mk m \over MT}x)}
\end{equation}
The term with the extra factor $\zeta_{Q}$ reflects the Pauli suppression of
the quark in the final state of the Compton process.

Again, if we want to get only the spectrum of invariant masses we should
use a function
\begin{eqnarray}
U_{c}(x,T/b)= \int {d^{3}{\bf k} \over 2k_{0}}e^{-k_{0}/T}C(x,T/b)=
2\pi M^{2} {K_{1}(Z) \over   Z },  \nonumber \\
 Z={1 \over T}\sqrt{M^{2}+m^{2}b^{2}+2mMbx_{0} },
\end{eqnarray}
which may be separately integrated over the history of the emitting plasma.

{\bf 6.} Having reached our goal to present
the expression for the rates of emission
in the form of simple quadratures we can easily find them numerically.
We can analyze the reliability of different analytic approximations and compare
them with those known in the literature.

At Fig.1 we present dependence (14) and (16) of the rate of emission of the
low mass dileptons as a function of  total momentum $\mk$ at a given
mass $M$. We consider the
annihilation (solid lines) and Compton (dashed lines)
channels separately and apart from the common
factor $10^{-9}\exp(-E/T)  \approx 10^{-9}\exp(-\mk /T)$.
The fugacities are taken
$\zeta_Q = \zeta_G =1 $. (The actual  values  of  the  fugacities  which
are the measure of the chemical equilibrium are about $\zeta_Q \sim 0.5$
and $\zeta_G \sim 0.75$ [1].)

 The limit of the very low masses, is of a special interest.
The formal limit of $M \rightarrow 0$ immediately reproduces
polarization loop contribution to the real photon emission [4],
\begin{eqnarray}
\pi_{ann}={ie^{2}g^{2}N_{c}C_{F} \over  2\pi^{3} } T^{2} e^{-ku/T}
\zeta_{Q}^{2} \int_{\xi}^{\infty} dy
[e^{-y} + {\zeta_{G} \over 2} e^{-2y} ]\times  \nonumber \\
\times [(1+{\xi\over y}-{\xi^{2} \over 2y^{2}})
 2\cosh^{-1}\sqrt{{y \over \xi }}+(1+{\xi \over y})\sqrt{1-{\xi\over y}}  ]
 \end{eqnarray}
\begin{eqnarray}
 \pi_{compt}=-{ie^{2}g^{2}N_{c}C_{F} \over  2\pi^{3} } e^{-ku/T} T^{2}
 \zeta_{Q}\zeta_{G} \int_{0}^{\infty} dy
 [e^{-(y+{\lambda^{2} \over 4y})}-
 {\zeta_{Q} \over 2} e^{-2(y+{\lambda^{2} \over 4y})} ] \times \nonumber \\
 \times [(1-{\xi \over y}-
 {\xi^{2} \over 2y^{2}} )\ln(1+{4y \over \xi}) +{2\xi \over y}+
 {4y(\xi+2y) \over (\xi +4y)^{2} }  ]
\end{eqnarray}
 where $\;\;\xi=m^{2}/(ku)T$, $\;\;\lambda= m/T$.
 For small but finite dilepton mass these
 equations are exact up to the terms of the order $\;\; M/m \sim M/gT<<1$.

   Because of the pre-factor from Eq.(1) which is strongly contributed by the
longitudinal mode and has a threshold at $M=2m_{l}$ we can not reproduce a
literal transition to the photon rate.
Moreover, for the very polarization tensor this transition is nonanalytic
and one needs a care when approaching this limit in numerical calculations.

 There is some mystery in regularization of the collinear singularity
 by means of the very dilepton mass [5]. Physically it would have meant that
the
 virtuality of the intermediate photon restricts the domain of coherence
 of quark-gluon and quark-photon interaction, which is hardly in line with
 the electromagnetic transparency of the quark-gluon plasma.

 Mathematically,  to achieve this regularization, the virtual mass and vertex
 radiative corrections were taken into account. But they unavoidably contain
the
trivial IR singularities at low gluon momenta (some of them with the extra
powers of fugacities) [2]. These should be cancelled out by the emission and
even absorption of the  real soft  gluons but for low-mass dileptons neither
the emitted gluons can be too soft nor they can be absorbed.

{\bf 7.Conclusion.}
In the limit of the  low masses,  the dilepton  rate of emission can be
expressed as a trace of the electromagnetic polarization tensor at $M=0$
(Eqs.(17) and (18)) times kinematic factor from the Eq.(1). In the
approximation of the leading logarithm and at $M>>2m_l$ it looks as
\begin{equation}
 {{dN_{e^{+}e^{-}}}\over{d^{4} k d^{4}x}}=
 {{5\alpha^{2}\alpha_{s}}\over{27\pi^{5}}}{T^2 \over M^{2}}
 (\zeta_Q +\zeta_G) e^{-ku/T}\ln{4(ku)T \over m_{therm}^{2}}.
\end{equation}
and there is no Born's term which accompany it (compare [5]).
In virtue of the analysis of the accuracy of this approximation [4]
which was performed for the real photons this formula essentially
underestimate the rate of emission. We recommend to use a simple
integral representations (11) and (13) instead of it.

We consciously do not integrate the rates over the history of the system
which can be easily done for a simple form of the hydrodynamic background.
The very approach of nonequilibrium field kinetics [2,3] was designed in
order to use the nonequilibrium partons distributions generated in a
"realistic" cascade.

 \vspace{1.5 cm}
I am indebted to G. Brown, E. Shuryak and the Nuclear Theory group at
SUNY at Stony Brook for continuous support.

 I am grateful to  E. Shuryak and I.Zahed for
many fruitful and helpful discussions.
\vskip 1.5cm

 \centerline{\bf REFERENCES}
  \vspace{1.0 cm}
 \ni 1. E. Shuryak: Phys. Rev. Lett. 68 (1992) 3270. \\
 \ni 2. A.Makhlin : Preprint SUNY-NTG-92-11.       \\
 \ni 3. A.Makhlin :  Preprint SUNY-NTG-93-20  \\
 \ni 4. A.Makhlin     Preprint SUNY-NTG-93-10     \\
 \ni 5. T. Altherr, P.V. Ruuskanen: Nucl.Phys. B380(1992)337 \\

\end{document}